\documentclass[submission,copyright,creativecommons]{eptcs}
\providecommand{\event}{LFMTP 2021} 
\usepackage{breakurl}             
\usepackage{underscore}           

\title{Adelfa: A System for Reasoning about LF Specifications}
\author{Mary Southern \qquad\qquad Gopalan Nadathur
\institute{University of Minnesota\\Minneapolis Minnesota, USA}
\email{south163@umn.edu \qquad\qquad ngopalan@umn.edu}
}
\def\titlerunning{A System for Reasoning about LF Specifications}
\def\authorrunning{M. Southern \& G. Nadathur}

\usepackage{amsmath, amsthm, amssymb, amsfonts}
\usepackage{proof}
\usepackage{stmaryrd}

\newcommand{\logic}{$\mathcal{L}_{LF}$}

\newcommand{\ie}{\mbox{\rm i.e.}}

\newcommand{\lfcheck}[4][\Sigma]{#2 \vdash_{#1} #3 \Leftarrow #4}

\newcommand{\seq}[5]{#1;#2;#3;#4\longrightarrow #5}
\newcommand{\seqsans}[3]{#1 ; \emptyset ; \emptyset ; #2\longrightarrow #3}
\newcommand{\ctxty}[2]{#1\lbrack {#2}\rbrack}
\newcommand{\ctxvarty}[3]{#1\!\uparrow\!#2 : #3}

\newcommand{\erase}[1]{({#1})^-}

\newcommand{\lflam}[2]{\lambda {#1}.{\mkern 3mu} {#2}}
\newcommand{\typedpi}[3]{\Pi {#1}{:}{#2}.{\mkern 3mu} #3}
\newcommand{\type}{\mbox{\it Type}}
\newcommand{\imp}[2]{#1\rightarrow #2}
\newcommand{\app}{\;}

\newcommand{\eq}{{\it eq}}
\newcommand{\refl}{{\it refl}}
\newcommand{\unit}{\mbox{\it unit}}
\newcommand{\sarr}{{\it arr}}
\newcommand{\sapp}{{\it app}}
\newcommand{\sabs}{{\it abs}}
\newcommand{\sty}{{\it ty}}
\newcommand{\stm}{{\it tm}}
\newcommand{\of}{{\it of}}
\newcommand{\ofapp}{{\it of\_app}}
\newcommand{\ofabs}{{\it of\_abs}}
\newcommand{\tyctx}{{\it c}}

\newcommand{\oty}{o}

\newcommand{\emptycs}{\cdot}
\newcommand{\emptybd}{\cdot}
\newcommand{\emptyce}{\cdot}
\newcommand{\fatm}[2]{\left\{#1 \vdash #2\right\}}
\newcommand{\fand}[2]{#1\wedge #2}
\newcommand{\for}[2]{#1\vee #2}
\newcommand{\fimp}[2]{#1\supset #2}
\newcommand{\fall}[2]{\forall #1. #2}
\newcommand{\fexists}[2]{\exists #1. #2}

\newcommand{\fctx}[3]{\Pi\,#1 : #2.#3}

\newcommand{\ftrue}{\top}
\newcommand{\ffalse}{\bot}

\newcommand{\ltannaux}[2]{{#2}^{*^{#1}}}
\newcommand{\eqannaux}[2]{{#2}^{@^{#1}}}

\newcommand{\subst}[2]{#2[#1]}
\newcommand{\hsubst}[2]{#2\llbracket {#1} \rrbracket} 

\newcommand{\lfstr}{\mbox{\sl LF-str}}

\newcommand{\intseq}[2]{#1 \longrightarrow #2}
\newcommand{\propty}{\mbox{\sl proptm}}
\newcommand{\toptm}{\mbox{\sl top}}
\newcommand{\imptm}{\mbox{\sl imp}}
\newcommand{\andtm}{\mbox{\sl and}}
\newcommand{\hypty}{\mbox{\sl hyp}}
\newcommand{\concty}{\mbox{\sl conc}}
\newcommand{\inittm}{\mbox{\sl init}}
\newcommand{\toprtm}{\mbox{\sl topR}}
\newcommand{\andltm}{\mbox{\sl andL}}
\newcommand{\andrtm}{\mbox{\sl andR}}
\newcommand{\impltm}{\mbox{\sl impL}}
\newcommand{\imprtm}{\mbox{\sl impR}}

\begin{document}
\maketitle

\begin{abstract}
We present a system called Adelfa that provides mechanized support
for reasoning about specifications developed in the Edinburgh Logical
Framework or LF.
Underlying Adelfa is a new logic named \logic.
Typing judgements in LF are represented by atomic formulas in
\logic\ and quantification is permitted over contexts and terms that 
appear in such formulas. 
Contexts, which constitute type assignments to uniquely named 
variables that are modelled using the technical device of
\emph{nominal constants}, are characterized in
\logic\ by \emph{context schemas} that describe their inductive structure. 
We present these formulas and an associated semantics before sketching
a proof system for constructing arguments that are sound with respect
to the semantics.
We then outline the realization of this proof system in Adelfa and
illustrate its use through a few example proof developments.
We conclude the paper by relating Adelfa to existing systems for
reasoning about LF specifications. 
\end{abstract}

\section{Introduction}\label{sec:introduction}

This paper describes a proof assistant called Adelfa that supports
reasoning about specifications written in the Edinburgh Logical
Framework, or LF~\cite{harper93jacm}. 
Adelfa is based on a logic called \logic\ whose atomic formulas
represent typing judgements in LF, and quantification is permitted over both
contexts and terms which appear in these judgements. 
Term quantification is qualified by simple types that identify the 
functional structure of terms while forgetting dependencies.
Context quantification is similarly qualified by schemas that capture
the way contexts evolve during typing derivations in LF.
The logic is complemented by a sequent-calculus based proof system
that supports the construction of arguments of validity for relevant
formulas.
In addition to interpreting the logical symbols, the rules of the
proof system embody an understanding of LF derivability; particular
rules encode a case analysis style reasoning on LF judgements, the
ability to reason inductively on the heights of such derivations, and
an understanding of LF metatheorems. 
Adelfa is a tactics style theorem prover that implements this proof
system: the development of a proof proceeds by invoking one of a set
of sound tactic commands towards partially solving one of the proof
obligations comprising the current proof state.

The rest of the paper is structured as follows. 
In Section~\ref{sec:logic} we briefly describe the underlying logic.
Section~\ref{sec:proofsystem} then outlines the associated proof
system. 
In Section~\ref{sec:adelfa} we describe Adelfa and illustrate its use
through a few examples. 
Section~\ref{sec:conclusion} concludes the paper by contrasting our
work with other approaches that have been explored for reasoning about
LF specifications.

\section{A Logic for Articulating Properties of LF Specifications}\label{sec:logic}

We limit ourselves here to a brief overview of \logic, leaving its
detailed presentation to other work~\cite{nadathur21arxiv,southern21phd}.  
As already noted, the atomic formulas in \logic\ represent typing
judgements in LF, a calculus with which we assume the reader to be
already familiar. 
We begin by summarizing aspects of LF that are pertinent to
discussions in this paper.
We then present the formulas of \logic\ and illustrate their use in
stating properties about LF derivability. 

\subsection{LF Basics and Hereditary Substitution}
 
The variant of LF that is used in \logic\ is that presented
in~\cite{harper07jfp}.
The only expressions permitted in this version, which is referred to
as \emph{canonical LF}, are those that are in $\beta$-normal form. 
Moreover, the typing rules ensure that all well-formed expressions are
also in $\eta$-long form.

\begin{figure}[h]
\begin{center}
\begin{tabular}{l}

\begin{tabular}{ll}
$\sty : \type$ \hspace{1in} & $\stm : \type$\\[3pt]
$\unit : \sty$ & $\sapp : \imp{\stm}{\imp{\stm}{\stm}}$\\
$\sarr : \imp{\sty}{\imp{\sty}{\sty}}$
   & $\sabs : \imp{\sty}{\imp{(\imp{\stm}{\stm})}{\stm}}$
\end{tabular}\\[25pt] 

\begin{tabular}{l}
$\of : \imp{\stm}{\imp{\sty}{\type}}$\\[3pt]
$\ofapp :\typedpi{M_1}{\stm}
                 {\typedpi{M_2}{\stm}
                          {\typedpi{T_1}{\sty}
                                   {\typedpi{T_2}{\sty}}}}$\\
\qquad$\of\app M_1\app (\sarr\app T_1\app T_2) \rightarrow 
            \of\app M_2\app T_1 \rightarrow 
            \of\app (\sapp\app M_1\app M_2)\app T_2$\\[5pt]
            
$\ofabs  : \typedpi{T_1}{\sty}
                   {\typedpi{T_2}{\sty}
                            {\typedpi{R}{\imp{\stm}{\stm}}{}}}$\\
\qquad$(\typedpi{x}{\stm}
                { \of\app x\app T_1 \rightarrow 
                      \of\app (R\app x)\app {\it T_2}) \rightarrow 
                      \of \app (\sabs \app T_1 \app  (\lflam{x}{R\app x})) 
                          \app (\sarr\app T_1\app T_2)}$
\end{tabular}
\end{tabular}
\end{center}
\vspace{-0.3cm}
\caption{An LF signature for typing in the simply typed $\lambda$-calculus} 
\label{fig:stlc-in-lf}
\end{figure}

Object systems are encoded in LF by describing a signature which
provides a means for representing the relevant constructs in the
system as well as the relations between them. 
For example, if our focus is on typing judgements concerning the simply
typed $\lambda$-calculus (STLC), then we might use the signature shown in
Figure~\ref{fig:stlc-in-lf}.
This specification begins by identifying the LF types $\sty$ and
$\stm$ and constructors for these types that serve to build
LF representations of STLC types and terms; note that object language
abstractions are represented using LF abstractions, following the idea
of higher-order abstract syntax. 
Dependencies in types are then exploited to encode the typing relation
between STLC terms and types.
Specifically, the signature defines the type-level constant $\of$
towards this end: if {\it T} and {\it Ty} are LF terms of type $\stm$ and
$\sty$, respectively, then $(\of\app {\it T}\app {\it Ty})$ is
intended to denote that the STLC expressions represented by {\it T} and
{\it Ty} stand in this relation.
The term-level constants $\ofapp$ and $\ofabs$ that are identified as
constructors for this type serve to encode the usual typing rules for
applications and abstractions. 
In the use of such a specification, a central focus is on LF typing 
judgements of the form $\lfcheck{\Gamma}{M}{A}$ where
$\Sigma$ represents a well-formed signature, $A$ represents a
well-formed type and $\Gamma$ represents a well-formed context that
assigns types to variables that may appear in $M$ and $A$.
Such a judgement is an assertion that $M$ is typeable at the type $A$
relative to the signature $\Sigma$ and a context $\Gamma$ that assigns
particular LF types to the variables that appear in $M$ and $A$.
Such judgements are meant to translate into meaningful statements about
the object systems encoded by the signature.
For example, if $\Sigma$ is the signature presented in
Figure~\ref{fig:stlc-in-lf} and $A$ is a type of the form $(\of\app
{\it T}\app {\it Ty})$, then the judgement corresponds to the
assertion that the STLC term represented by ${\it T}$ is typeable at the
type represented by ${\it Ty}$ and that $M$ encodes an STLC derivation
for this judgement.  

The typing rules in LF require the consideration of substitutions into
LF expressions.
In canonical LF, substitution must build in normalization, a
requirement that is realized in~\cite{harper07jfp} via the operation
of \emph{hereditary substitution}.
We have generalized this operation to permit multiple simultaneous
substitutions.
Formally, a substitution is given by a finite set of triples of the
form
$\{\langle x_1,M_1,\alpha_1\rangle,\ldots,
   \langle x_n,M_n,\alpha_n\rangle\}$
in which, for $1 \leq i \leq n$, $x_i$ is 
a variable, $M_i$ is a term and $\alpha_i$ is a simple type
constructed from the sole base type $\oty$.
We refer to the simple types that index substitutions as \emph{arity
  types}.
The attempt to apply such a substitution to an LF expression always 
terminates.
Moreover, it must yield a unique result whenever the expression being
substituted into and the substituents satisfy the functional structure
determined by the arity types.
We do not discuss this constraint in more detail here, noting only
that the constraint is satisfied in all the uses we make of hereditary
substitution in this paper. 
We write $\hsubst{\theta}{E}$ to denote the result of applying the
substitution $\theta$ to the expression $E$ in this situation. 

\subsection{Formulas over LF Typing Judgements and their Meaning}
\label{ssec:formulas}

The logic \logic\ is parameterized by an LF signature $\Sigma$.
The basic building blocks for formulas in this context are typing
judgements of the form $\lfcheck{\Gamma}{M}{A}$ that are written as
$\fatm{G}{M : A}$.
However, the syntax of the expressions in the logic differs somewhat
from that of the LF expressions.
To begin with, we permit term variables that are bound by quantifiers
to appear in types and terms.
These variables have a different logical character from the variables
that are bound in a context.
More specifically, these variables may be instantiated by terms in the
domains of the quantifiers, whereas the variables bound by declarations
in an LF context represent fixed entities that are also distinct from
all other similar entities within the typing judgement. 
To accurately capture the role of the variables bound in an LF
context, we represent them using \emph{nominal
  constants}~\cite{gacek11ic,tiu06lfmtp}; these are entities that are
represented in this paper by the symbol $n$ possibly with subscripts and that
behave like constants except that they can be permuted with
other such entities in an atomic formula without changing logical content. 
To support this treatment, we also allow nominal constants to appear
in expressions corresponding to types and terms in the logic.
Finally, we allow for contexts to be variables so as to permit
quantification over them.
More specifically the syntax for $G$, which constitutes a \emph{context
expression} in $\fatm{G}{M : A}$, is given by the following rule:
\[
G \quad ::= \quad \Gamma\ |\ \cdot\ |\ G,n:A.
\]
The symbol $\Gamma$ here denotes the category of variables that range
over contexts. 

In typical reasoning scenarios, instantiations for context variables
must be constrained in order to be able to articulate contentful
properties. 
Such constraints are described via a special kind of typing
for context variables. 
The formal realization of this idea, which is inspired by the notion
of \emph{regular worlds} in Twelf~\cite{Pfenning02guide,schurmann00phd}, is
based on declarations given by the following syntax rules. 
  \[\begin{array}{rrcl}
\mbox{\bf Block Declarations} & \Delta & ::= & \emptybd\ \vert\ \Delta, y : A \\
\mbox{\bf Block Schema}   & \mathcal{B} & ::= & \{x_1:\alpha_1,\ldots, x_n:\alpha_n\}\Delta\\
\mbox{\bf Context Schema} & \mathcal{C} & ::= & \emptycs\ \vert\ \mathcal{C}, \mathcal{B}
\end{array}\]
According to this definition, a \emph{context schema} is a collection of
block schemas. 
The instances of a block schema 
$\{x_1:\alpha_1,\ldots,x_n:\alpha_n\}(y_1:B_1,\ldots,y_m:B_m)$
are sequences $n_1:C_1,\ldots,n_m:C_m$ obtained by choosing
particular terms for the schematic variables $x_1,\ldots,x_n$ and particular
nominal constants for the $y_1,\ldots,y_m$.
A context satisfies a context schema if the context comprises
a sequence of instances for the block schemas defining it.
Context schemas are the types that are associated with context
variables and they play the obvious role in limiting their domains.

The formulas of \logic\ are given by the following syntax rule:
\[
F ::= \fatm{G}{M : A}\ |\ \fall{x:\alpha}{F}\ |\ \fexists{x:\alpha}{F}\ |\ \fctx{\Gamma}{\mathcal{C}}{F}
      \ |\ \fand{F_1}{F_2}\ |\ \for{F_1}{F_2}\ |\ \fimp{F_1}{F_2}\ |\ \ftrue\ |\ \ffalse
\]
The symbol $\Pi$ represents universal quantification pertaining to
contexts; note that such quantification is qualified by a context
schema.
The symbol $x$ represents a term variable, i.e. the logic permits
universal and existential quantification over LF terms.
The symbol $\alpha$ that annotates such variables represents an arity
type.
For a formula to be well-formed, the terms appearing in it must be
well-typed and in canonical form relative to an arity typing that
is determined as follows: for the quantified variables this is given
by their annotations; and for a constant in the signature or a nominal
constant in the context expression that is assigned the LF type $A$ it
is the \emph{erased form} of $A$, written as $\erase{A}$, that is
obtained by replacing the atomic types in $A$ by $\oty$ and otherwise
retaining the functional structure.
This well-typing requirement is intended to eliminate structural
issues in the consideration of validity of a 
formula, thereby allowing the focus to be on the more contentful
relational aspects that are manifest in dependencies in typing. 

A closed formula, i.e. a formula in which no unbound context or term
variable appears, that is of the form $\fatm{G}{M : A}$ is deemed to
be true exactly when $G$ is a well-formed LF context, $A$ is a well
formed LF type relative to $G$ and the typing judgement
$\lfcheck{G}{M}{A}$ is derivable.  
This semantics is extended to closed formulas involving the logical
connectives and constants by using their usual understanding.
Finally, the quantifiers are accorded a substitution semantics.
The formula $\fctx{\Gamma}{\mathcal{C}}{F}$ holds just in the case that
$\subst{\{G/\Gamma\}}{F}$ holds for every context $G$ that satisfies the 
context schema $\mathcal{C}$, where $\subst{\sigma}{E}$ denotes the result of
a standard replacement substitution applied to the expression $E$,
being careful, of course, to carry out any renaming of bound variables
that is necessary to avoid inadvertent capture.
The formula $\fall{x:\alpha}{F}$ holds exactly when 
$\hsubst{\{\langle x, t, \alpha\rangle\}}{F}$ holds
for every closed term $t$ that has the arity type 
$\alpha$.
Finally, the formula $\fexists{x:\alpha}{F}$ is true if there is some
closed term $t$ with arity type $\alpha$ such that
$\hsubst{\{\langle x, t, \alpha\rangle\}}{F}$ is true.

\subsection{An Illustration of the Logic}\label{ssec:example}

The property of uniqueness of type assignment for the STLC
can be expressed via the formula
\begin{tabbing}
\hspace{1cm}\=\hspace{3cm}\=\kill
\>$\fctx{\Gamma}{\tyctx}{\fall{E:\oty}{\fall{T_1:\oty}{\fall{T_2:\oty}{\fall{D_1:\oty}{\fall{D_2:\oty}{\fimp{\fatm{\Gamma}{T_1:\sty}}{\fimp{\fatm{\Gamma}{T_2:\sty}}{}}}}}}}}$\\
\>\>$\fimp{\fatm{\Gamma}{D_1:\of\app E\app T_1}}{\fimp{\fatm{\Gamma}{D_2:\of\app E\app T_2}}{\fexists{D_3:\oty}{\fatm{\emptyce}{D_3:\eq\app T_1\app T_2}}}}$
\end{tabbing}
where $\tyctx$ represents the context schema
$\{T:\oty\}(x:tm,y:\of\app x\app T)$ and the signature parameterizing the
logic is that in Figure~\ref{fig:stlc-in-lf} augmented with the
declarations $\eq : ty \rightarrow ty \rightarrow \type$ and
$\refl : \Pi T:ty. \eq\app T\app T$.
A special case of this formula is that when the context variable is
instantiated with the empty context.
However, the typing rule for abstractions will require us to consider
non-empty contexts in the analysis.
Note, though, that the signature ensures that the extended contexts
that need to be considered will always satisfy the constraint
expressed by $\tyctx$.

We can argue for the validity of the formula above in two steps.
We show first the validity of 
\begin{tabbing}
\hspace{2.6cm}\=\kill
\>$\fctx{\Gamma}{\tyctx}
       {\fall{T_1:\oty}
             {\fall{T_2:\oty}
                   {\fall{D:\oty}
                         {\fimp{\fatm{\Gamma}{D:\eq\app T_1\app T_2}}
                               {\fatm{\emptyce}{D:\eq\app T_1\app T_2}}}}}}.$
\end{tabbing}
This formula, which is essentially a \emph{strengthening} property
for $\eq$, can be argued to be valid by observing that bindings in a
well-formed LF context satisfying the context schema $\tyctx$ have no
role to play in constructing a term of the (well-formed) LF type
$(\eq\app T_1\app T_2)$ for any terms $T_1$ and $T_2$.
We then combine this observation with the validity of the formula
\begin{tabbing}
\hspace{1cm}\=\hspace{3cm}\=\kill
\>$\fctx{\Gamma}{\tyctx}{\fall{E:\oty}{\fall{T_1:\oty}{\fall{T_2:\oty}{\fall{D_1:\oty}{\fall{D_2:\oty}{\fimp{\fatm{\Gamma}{T_1:\sty}}{\fimp{\fatm{\Gamma}{T_2:\sty}}{}}}}}}}}$\\
\>\>$\fimp{\fatm{\Gamma}{D_1:\of\app E\app T_1}}{\fimp{\fatm{\Gamma}{D_2:\of\app E\app T_2}}{\fexists{D_3:\oty}{\fatm{\Gamma}{D_3:\eq\app T_1\app T_2}}}}$.
\end{tabbing}

To establish the validity of the last formula, we must show that, for
a closed context expression $G$ that instantiates the schema $\tyctx$
and for closed expressions $E$, $T_1$, $T_2$, $D_1$, and $D_2$, if
$\fatm{G}{D_1:\of\app E\app T_1}$ and $\fatm{G}{D_2:\of\app E\app
  T_2}$ are both valid, then $T_1$ and $T_2$ must be identical.
%
The argument proceeds by induction on the height of the derivation of
$\lfcheck{G}{D_1}{\of\app E\app T_1}$ that the assumption implies must
exist. 
An analysis of this derivation shows that there are three cases to
consider, corresponding to whether the head symbol of
$D_1$ is $\ofapp$, $\ofabs$, or a nominal constant from $G$.
For the last of these, the argument is that the validity of $\fatm{G}{D_1:\of\app E\app T_1}$
implies that $G$ is a well-formed context and thus that the typing is unique.
The other two cases invoke the induction hypothesis; in the case of $\ofabs$
we must consider a shorter derivation in which the context has been
extended, but in a way that conforms to the definition of the context
schema $\tyctx$. 

\section{A Proof System for the Logic}\label{sec:proofsystem}

We now describe a sequent calculus that supports arguments of validity
of the kind outlined in Section~\ref{ssec:example}.
We aim only to present the spirit of this calculus, leaving
its detailed consideration again to other
work~\cite{nadathur21arxiv,southern21phd}.
We note also that the main goal for the calculus is to provide
a means for sound and effective reasoning rather than to be complete
with respect to the semantics described for \logic.
We begin by explaining the structure of sequents before proceeding to
a discussion of the inference rules.

\subsection{The Structure of Sequents}\label{ssec:sequents}

A sequent, written as $\seq{\mathbb{N}}{\Psi}{\Xi}{\Omega}{F}$, is a
judgement that relates a finite subset $\mathbb{N}$ of nominal
constants with associated arity types, a finite set $\Psi$ of term
variables also with associated arity types, a finite set $\Xi$ of
context variables with types of the kind described below, a finite set
$\Omega$ of \emph{assumption formulas} and a \emph{conclusion} or
\emph{goal} formula $F$; 
here, $\mathbb{N}$ is the \emph{support set} of the
sequent, $\Psi$ is its \emph{eigenvariables context} and $\Xi$ is its
\emph{context variables context}.
The formulas in $\Omega \cup \{F\}$ must be formed out of the symbols
in $\mathbb{N}$, $\Psi$, $\Xi$ and the (implicit) signature $\Sigma$,
and they must be well-formed with respect to these collections in the
sense explained in Section~\ref{ssec:formulas}. 
The members of $\Xi$ have the form
$\ctxvarty{\Gamma}{\mathbb{N}_{\Gamma}}{\ctxty{\mathcal{C}}{G_1,\ldots,G_n}}$,
where $\mathbb{N}_{\Gamma}$ is a collection of nominal constants,
$\mathcal{C}$ is a context schema, and $G_1,\ldots,G_n$ is a listing of
instances of block schemas from $\mathcal{C}$ in which the types assigned to
nominal constants are well-formed with respect to $\Sigma$,
$\mathbb{N} \setminus \mathbb{N}_\Gamma$, and $\Xi$. 
This ``typing'' of the variable $\Gamma$ is intended to limit its
range to closed contexts obtained by interspersing instances of block
schemas from $\mathcal{C}$ in which nominal constants from
$\mathbb{N}_\Gamma$ do not appear between instances of
$G_1,\ldots,G_n$ obtained by substituting terms formed from $\Sigma$
and nominal constants not appearing in the support set of the
sequent for the variables in $\Psi$; the substitutions for the
variables in $\Psi$ must respect arity typing and the LF types in the
resulting context must be well-typed in an arity sense. 

The basic notion of meaning for sequents is one that pertains to
closed sequents, i.e. ones of the form
$\seqsans{\mathbb{N}}{\Omega}{F}$. 
Such a sequent is \emph{valid} if $F$ is valid or one of the
assumption formulas in $\Omega$ is not valid.
A sequent of the general form
$\seq{\mathbb{N}}{\Psi}{\Xi}{\Omega}{F}$ is then considered valid if
all of its instances obtained by substituting closed terms not
containing the nominal constants in $\mathbb{N}$ and respecting arity
typing constraints for the variables in $\Psi$ and replacing the
variables in $\Xi$ with closed contexts respecting their types in the
manner described above are valid.
The goal of showing that a formula $F$ whose nominal
constants are contained in the set $\mathbb{N}$ is valid now reduces to
showing the validity of the sequent $\seq{\mathbb{N}}{\emptyset}{\emptyset}{\emptyset}{F}$.

\subsection{The Rules for Deriving Sequents}\label{ssec:proofrules}

The sequent calculus comprises two kinds of rules: those that pertain
to the logical symbols and structural aspects of sequents and those
that encode the interpretation of atomic formulas as assertions of
derivability in LF. We discuss the rules under these categories below,
focusing mainly on the latter kind of rules. For paucity of space, we
do not present the rules explicitly but rather discuss their intuitive
content. We also note that all the rules that we describe have been
shown to be sound~\cite{nadathur21arxiv}.

\subsubsection{Structural and Logical Rules}

The calculus includes the usual contraction and weakening rules
pertaining to assumption formulas.
Also included are rules for adding and removing entries from the
support set and the eigenvariables and context variables contexts when
these additions do not impact the overall well-formedness of
sequents.
Finally, the cut rule, which facilitates the use of well-formed
formulas as lemmas, is also present in the collection.

The most basic logical rule is that of an axiom.
The main deviation from the usual form for this is the
incorporation of the invariance of LF derivability under permutations
of the names of context variables; this is realized in our proof
system via a notion of equivalence of formulas under permutations of nominal
constants.   
The rules for the connectives and quantifiers take the expected form.
The only significant point to note is that eigenvariables that are
introduced for (essential) universal quantifiers must be
raised over the support set of the sequent to correctly reflect
dependencies given our interpretation of sequents~\cite{miller92jsc}.

\subsubsection{The Treatment of Atomic Formulas}

The calculus builds in the understanding of formulas of the form
$\fatm{G}{M : A}$ via LF derivability.
If $A$ is a type of the form $\typedpi{x}{A_1}{A_2}$, then $M$ must
have the form $\lflam{x}{M'}$ and the atomic formula can be replaced
by one of the form $\fatm{G,n:A_1}{M':A_2}$ in the sequent; here, $n$
must be a nominal constant not already in the support set and if $G$
contains a context variable then its type annotation must be changed
to prohibit the occurrence of $n$ in its instantiations.
If $A$ is an atomic type and $\fatm{G}{M : A}$ is the goal formula,
then the corresponding rule allows a step to be taken in the
validation of the typing judgement; specifically, if $M$ is the term
$(h\app M_1\app\ldots\app M_n)$ where $h$ is a constant or a nominal
constant to which $\Sigma$ or $G$ assigns the
LF type $\typedpi{x_1}{A_1}{\ldots\typedpi{x_n}{A_n}{A'}}$ and $A$ is
identical to $\hsubst{\{\langle x_1,M_1\erase{A_1} \rangle,
  \ldots,\langle x_n,M_n,\erase{A_n}\rangle\}}{A'}$, then the rule
leads to the consideration of the derivation of sequents in which
the goal formula is changed to $\fatm{G}{\hsubst{\{\langle x_1,M_1\erase{A_1} \rangle,
  \ldots,\langle x_{i-1},M_{i-1},\erase{A_{i-1}}\rangle\}}{M_i}}$ for
$1 \leq i \leq n$. Note that if $G$ begins with a context variable $\Gamma$,
  then the assignments in the blocks in the ``type'' of $\Gamma$ are
  considered to be assignments in $G$.

The most contentful part of the treatment of atomic formulas
is when the formula $\fatm{G}{M : A}$ in which $A$ is an atomic type
appears as an assumption formula in the sequent. 
The example in Section~\ref{ssec:example} demonstrates the
\emph{case analysis} style of reasoning that we would want to capture 
in the proof rule; we must identify all the possibilities for the valid
closed instances of this formula and analyze the validity of the
sequent based on these instances.
The difficulty, however, is that there may be far too many closed
instances to consider explicitly.
This issue can be refined into two specific problems that must be
addressed. 
First, the context $G$ might begin with a context variable and we must
then identify a realistic way to consider all the instantiations of that
variable that yield an actual, closed context.
Second, we must describe a manageable approach to considering all
possible instantiations for the term variables that may appear in
$\fatm{G}{M : A}$. 

The first problem is solved in the enunciation of the rule through an
\emph{incremental elaboration} of a context variable that is driven by
the atomic formula under scrutiny.
Suppose that $G$ begins with the context variable $\Gamma$
corresponding to which there is the declaration
$\ctxvarty{\Gamma}{\mathbb{N}_{\Gamma}}{\ctxty{\mathcal{C}}{G_1,\ldots,G_k}}$
in the context variables context.
We would at the outset need to consider all the constants in $\Sigma$
and all the nominal constants identified explicitly in $G$, which
includes the ones declared in $G_1,\ldots,G_k$, as potential heads for
$M$ in the formula $\fatm{G}{M:A}$.
Additionally, this head may come from a part of $\Gamma$ that has not
yet been made explicit.
To account for this, the rule considers all the possible instances for
the block declarations constituting $\mathcal{C}$ and all possible
locations for such blocks in the sequence $G_1,\ldots,G_k$.
We note that the number of such instances that have to be examined 
is finite because it suffices to consider exactly one representative
for any nominal constant that does not appear in the support set of
the sequent; this observation follows from the invariance of LF typing
judgements under permutations of names for the variables bound in the
context. 

The second problem is addressed by first describing a notion
of unification that will ensure that all closed instances will be
considered and then identifying the idea of a \emph{covering set of
unifiers} that enables us to avoid an exhaustive consideration.
To elaborate a little on this approach, suppose that the (nominal)
constant $h$ with LF type
$\typedpi{x_1}{A_1}{\ldots\typedpi{x_n}{A_n}{A'}}$ has been identified 
as the candidate head for $M$.
Further, for $1\leq i \leq n$, let $t_i$ be terms representing fresh
variables raised over the support set of the sequent.\footnote{If
  $n_1,\ldots,n_{\ell}$ is a listing of $\mathbb{N}$ then, for $1 \leq
  i \leq n$, $t_i$ is the term $(z_i\app n_1\app \cdots\app n_{\ell})$
  where $z_i$ is a fresh variable of suitable arity type.}
Then, based on the notion of unification described, for each
substitution $\theta$ that unifies 
$(h\app t_1\app\ldots\app t_n)$ and $M$ on the one hand and $A$ and
$\hsubst{\{\langle x_1,t_1\erase{A_1} \rangle,
  \ldots,\langle x_n,t_n,\erase{A_n}\rangle\}}{A'}$ on the other, 
it suffices to consider the derivability of the sequent that results
from replacing $\fatm{G}{M:A}$ in the original sequent with the set of
formulas 
\[
  \{\fatm{G}{t_i : \hsubst{\{\langle x_1,t_1\erase{A_1} \rangle,
  \ldots,\langle
  x_{i-1},t_{i-1},\erase{A_{i-1}}\rangle\}}{A_i}}\ \vert\ 1 \leq i \leq
  n\}
\]
and then applying the substitution $\theta$.
However, this will still result in a a large number of cases since the
collection of unifiers must include all relevant closed instances for
the analysis to be sound.
The notion of a covering set provides a means for limiting attention
to a small subset of unifiers while still preserving soundness.

\subsubsection{Induction over the Heights of LF Derivations}

The example in Subsection~\ref{ssec:example} also illustrates the role
of induction over the heights of LF typing derivations in informal
reasoning. 
This kind of induction is realized in our sequent calculus by using an
annotation based scheme inspired by Abella~\cite{baelde14jfr,
  gacek09phd}.
Specifically, we add to the syntax two additional forms of atomic
formulas: $\eqannaux{i}{\fatm{G}{M:A}}$ and
$\ltannaux{i}{\fatm{G}{M:A}}$.
These represent, respectively, a formula that has an LF derivation of
some given height and another formula of strictly smaller height; the
latter formula is obtained typically by an unfolding step embodied in
the use of a case analysis rule.
The index $i$ on the annotation symbol is used to identify distinct pairs of 
$@$ and $*$ annotations.
The induction proof rule then has the form

\[
\infer[induction]
      {\seq{\mathbb{N}}
           {\Psi}
           {\Xi}
           {\Omega}
           {\mathcal{Q}_1.(
             \fimp{F_1}{\fimp{\ldots}
                       {\mathcal{Q}_{k-1}.(
                         \fimp{F_{k-1}}
                              {\mathcal{Q}_k.
                                  (\fimp{\fatm{G}{M:A}}
                                        {\fimp{\ldots}{F_n}})})}})}}
      {\deduce{\mathcal{Q}_1.( 
                  \fimp{F_1}{\fimp{\ldots}
                                  {\mathcal{Q}_{k-1}}.(
                                    \fimp{F_{k-1}}
                                         {\mathcal{Q}_k.
                                            (\fimp{\eqannaux{i}{\fatm{G}{M:A}}}
                                                  {\fimp{\ldots}{F_n}})})})}
              {\seq{\mathbb{N}}
                   {\Psi}
                   {\Xi}
                   {\Omega,
                           \mathcal{Q}_1.( 
                              \fimp{F_1}{\fimp{\ldots}
                                              {\mathcal{Q}_{k-1}}.(
                                               \fimp{F_{k-1}}
                                                    {\mathcal{Q}_k.
                                                      (\fimp{\ltannaux{i}{\fatm{G}{M:A}}}
                                                            {\fimp{\ldots}{F_n}})})})}
                   {}}}
\]
where $\mathcal{Q}_i$ represent a sequence of context quantifiers or universal
term quantifiers and the annotations $@^i$ and $*^i$ must not already appear
in the conclusion sequent.
The premise of this proof rule can be viewed as providing a proof schema for
constructing an argument of validity for any particular height $m$, and so by 
an inductive argument we can conclude that the formula will be valid 
regardless of the derivation height.
This idea is made precise in a proof of soundness for the
rule~\cite{nadathur21arxiv,southern21phd}.

For this proof rule to be useful in reasoning, we will need a form of
case analysis which permits us to move from a formula annotated with
$@$ to one annotated by $*$ when reduced. 
Such a mechanism is built into the proof system and is used in the
examples in the next section.

\subsubsection{Rules Encoding Metatheorems Concerning LF Derivability}

Typing judgements in LF admit several metatheorems that are useful in
reasoning about specifications: if $\lfcheck{\Gamma}{M}{A}$ has a
derivation then so does $\lfcheck{\Gamma, x : A'}{M}{A}$ for a fresh
variable $x$ and any well-formed type $A'$ (weakening); if
$\lfcheck{\Gamma, x : A'}{M}{A}$ has a derivation and $x$ does not
appear in $M$ or $A$ then so also does $\lfcheck{\Gamma}{M}{A}$
(strengthening); if $\lfcheck{\Gamma_1, x_1 : A_1, x_2 : A_2, \Gamma_2}{M}{A}$ has
a derivation and $x_1$ does not appear in $A_2$ then 
$\lfcheck{\Gamma_1, x_2 : A_1, x_1 : A_1, \Gamma_2}{M}{A}$ must have a derivation
(permutation); and if $\lfcheck{\Gamma_1}{M'}{A'}$ and
$\lfcheck{\Gamma_1,x:A',\Gamma_2}{M}{A}$ have derivations then there
must be a derivation for
$\lfcheck{\Gamma_1,\hsubst{\{\langle x,M',\erase{A'}\rangle\}}{\Gamma_2}}
         {\hsubst{\{\langle x,M',\erase{A'}\rangle\}}{M}}
         {\hsubst{\{\langle x,M',\erase{A'}\rangle\}}{A}}$ (substitution). 
Moreover, in the first three cases, the derivations are structurally
similar, e.g. they have the same heights.
These metatheorems are built into the sequent calculus via (sound)
axioms.
For example, (one version of) the strengthening metatheorem is encoded
in the axiom
\[
\infer[\lfstr]
      {\seq{\mathbb{N}}{\Psi}{\Xi}{\Omega}{\fimp{\fatm{G,n:B}{M:A}}{\fatm{G}{M:A}}}}
      {\begin{array}{c}
         n\mbox{ does not appear in }M\mbox{, } A\mbox{, or the explicit bindings in }G
       \end{array}}
\]
These axioms can be combined with the cut rule to encode the informal reasoning process. 

\section{The Adelfa System and its Use in Reasoning}\label{sec:adelfa}
 
In this section we expose the Adelfa system that provides support for
mechanizing arguments of validity for formulas in \logic\ 
using the proof system outlined in
Section~\ref{sec:proofsystem}.
The first subsection below provides an overview of Adelfa.
We then consider a few example reasoning tasks to indicate how the
system is intended to be used and also to give a sense for its
capabilities.
 
\subsection{Overview of Adelfa}
The structure of Adelfa is inspired by the proof assistant 
Abella~\cite{baelde14jfr,gacek09phd}.
Each development in Adelfa is parameterized by an LF signature which
is identified by an initial declaration. 
The interaction then proceeds to a mode in which context schemas can
be identified and theorems can be posited.
When a theorem has been presented, the interaction enters a
\emph{proof mode} in which the user directs the construction of a
proof for a suitable sequent using a repertoire of tactics.
These tactics encode the (sound) application of a combination of rules
from the sequent calculus to produce a new proof state represented by
a collection of sequents for which proofs need to be constructed.
We note in this context the presence of a {\it search} tactic that
attempts to construct a proof for a sequent with an atomic
goal formula through the repeated use of LF typing rules.
If a tactic yields more than one proof obligation then these are
ordered in a predetermined way and all existing obligations are
maintained in a stack, to be eventually treated in order for the proof
to be completed. 

The syntax of LF terms and types in Adelfa follows that of fully
explicit Twelf.
The type $\typedpi{x}{A}{B}$ is represented by $\{x:A\}~B$ and the
abstraction term $\lflam{x}{M}$ is represented by $\lbrack x\rbrack~M$.
A context schema definition is introduced by the keyword {\it Schema}
and it associates an identifier with the schema that is to be used in
its place in the subsequent development. 
The block schemas defining a context schema are separated by $;$, the 
schematic variable declarations of each block schema are presented
within braces, and the block itself is surrounded by parenthesis.
A theorem is introduced by the keyword {\it Theorem}, and the
declaration associates an identifier with the formula to be proved. 
Once a theorem has been successfully proved this identifier is available for
use as a lemma (via the cut rule) in later proofs.

The following table identifies the formula syntax of Adelfa.

\begin{center}
\begin{tabular}{r|lcr|l}
Formula & Adelfa Syntax & \hspace{1in} & Formula & Adelfa Syntax\\
\hline
$\fall{x:\alpha}{F}$ & $\mbox{forall }x:\alpha.~F$ & &
$\for{F_1}{F_2}$ & $F_1~\backslash/~F_2$\\
$\fexists{x:\alpha}{F}$ & $\mbox{exists }x:\alpha.~F$ & &
$\fimp{F_1}{F_2}$ & $F_1~\mbox{=\textgreater}~F_2$\\
$\fctx{G}{\mathcal{C}}{F}$ & $\mbox{ctx }G:\mathcal{C}.~F$ & &
$\ftrue$ & true\\
$\fand{F_1}{F_2}$ & $F_1~/\backslash~F_2$ & & $\ffalse$ & false
\end{tabular}
\end{center}

A key issue in the Adelfa implementation is the realization of case analysis.
In Section~\ref{sec:proofsystem}, we have discussed the basis for this
rule in a special form of unification and also the role of
the idea of covering sets of unifiers in its practical realization.
We have shown that, in the situations where it is applicable, the
notion of higher-order pattern unification~\cite{miller91jlc} can be
adapted to provide such covering sets of solutions~\cite{southern21phd}.
Adelfa employs this approach, using the higher-order pattern
unification algorithm described by Nadathur and
Linnell~\cite{nadathur05iclp} in its implementation.
The benefit of following this course is that the covering set of
solutions comprises a single substitution. 

\subsection{Example Developments in Adelfa}
We begin by discussing the formalization of the proof of type
uniqueness for the STLC in Adelfa, thereby demonstrating the use of
both induction and case analysis in reasoning.
We then show the usefulness of LF metatheorems by considering the
proof of cut admissibility for a simple sequent calculus. 
Finally, we demonstrate the flexibility of the logic through a simple example
which is of a form that cannot be directly represented as a function type; we
will see in the next section that the ability to reason directly about such 
formulas distinguishes Adelfa from the Twelf family of systems.

\subsubsection{Type Uniqueness for the STLC}

The informal argument in Section~\ref{ssec:example} had identified a
context schema that is relevant to this example.
This schema would be presented to Adelfa via the following
declaration:
\begin{verbatim}
Schema c := {T:o}(x:tm,y:of x T).
\end{verbatim}
The proof of the actual type uniqueness theorem had relied on a
strengthening lemma concerning the equality relation between encodings
of types in the STLC.
We elide the proof of this lemma and its use in the formalization,
focusing instead on the proof of the formula
\begin{tabbing}
\hspace{1cm}\=\hspace{3cm}\=\kill
\>$\fctx{\Gamma}{\tyctx}{\fall{E:\oty}{\fall{T_1:\oty}{\fall{T_2:\oty}{\fall{D_1:\oty}{\fall{D_2:\oty}{\fimp{\fatm{\Gamma}{T_1:\sty}}{\fimp{\fatm{\Gamma}{T_2:\sty}}{}}}}}}}}$\\
\>\>$\fimp{\fatm{\Gamma}{D_1:\of~E~T_1}}{\fimp{\fatm{\Gamma}{D_2:\of~E~T_2}}{\fexists{D_3:\oty}{\fatm{\Gamma}{D_3:\eq\app
        T_1\app T_2}}}}$;
\end{tabbing}
the full development of this example and the others in this paper are available
with the Adelfa source from the Adelfa web page~\cite{adelfa.website}.

The proof development process starts with the presentation of the
above formula as a theorem.
This will result in Adelfa displaying the following proof state to the
user:

\begin{verbatim}
Vars:
Nominals: 
Contexts:
==================================
ctx G:c, forall E:o T1:o T2:o D1:o D2:o, 
  {G |- T1 : ty} => {G |- T2 : ty} => {G |- D1 : of E T1} => 
  {G |- D2 : of E T2} => exists d3:o, {G |- D3 : eq T1 T2}
\end{verbatim}

\noindent This proof state corresponds transparently to the sequent whose proof
will establish the validity of the formula; the components of the
sequent to the left of the arrow appear above the double line
and the goal formula appears below the double line.
To make it easy to reference particular formulas during reasoning, a
name is associated with each assumption formula in the sequent.
In the general case, where there can be more than one proof
obligation, the remaining obligations will be indicated by a listing
of just their goal formulas below the obligation currently in focus.

The informal argument used a induction on the height of the LF
derivation represented by the third atomic formula that appears in the
goal formula shown.
To initiate this process in Adelfa, we would invoke an induction
tactic command, indicating the atomic formula that is to be the
focus of the induction. 
This causes the goal formula with the third atomic formula annotated
with a \verb+*+ to be added to the assumption formulas and the goal
formula to be changed to one in which the third formula has the
annotation \verb+@+. 
At this stage, we may invoke a tactic command to apply a sequence of
right introduction rules, followd by another command to invoke case
analysis on the assumption formula
\verb+{G |- D1 : of E T1}@+.
Doing so results in the following proof state:
\begin{verbatim}
Vars: a1:o -> o -> o, R:o -> o, T3:o, T4:o, D2:o, T2:o
Nominals: n2:o, n1:o, n:o
Contexts: G{n2, n1, n}:c[]
IH: ctx G:c, forall E:o T1:o T2:o D1:o D2:o, 
      {G |- T1 : ty} => {G |- T2 : ty} => {G |- D1 : of E T1}* =>
      {G |- D2 : of E T2} => exists D3:o, {G |- D3 : eq T1 T2}
H1:{G |- arr T3 T4 : ty}
H2:{G |- T2 : ty}
H4:{G |- D2 : of (abs T3 ([x]R x)) T2}
H5:{G |- T3 : ty}*
H6:{G |- T4 : ty}*
H7:{G, n:tm |- R n : tm}*
H8:{G, n1:tm, n2:of n1 T3 |- a1 n1 n2 : of (R n1) T4}*

==================================
exists D3, {G |- D3 : eq (arr T3 T4) T2}

Subgoal 2: exists D3:o, {G |- D3 : eq T1 T2}
Subgoal 3: exists D3:o, {G |- D3 : eq (T1 n n1) (T2 n n1)}
\end{verbatim}

\noindent As is to be anticipated, case analysis results in the consideration of
three different possibilities: that where the head of \verb+D1+ is, respectively, 
the constant $\ofabs$, the constant $\ofapp$, or a
nominal constant from the context \verb+G+.
The first of these cases is shown in elaborated form, the other two
become additional proof obligations.
The analysis based on the first case replaces the original assumption
formula with new ones according to the type associated with \ofabs.
The annotation on these formulas is changed from \verb+@+ to \verb+*+,
to represent the fact that the derivations of the LF judgements
associated with them must be of smaller height.
Finally, those of these formulas that represent typing judgements for
(LF) abstractions are analyzed further, resulting in the introduction
of the new nominal constants \verb+n+, \verb+n1+, and \verb+n2+ into
the corresponding assumption formulas, and an annotation of the
context variable G that indicates that only those instantiations in
which these nominal constants do not appear must be considered for it.

At this point, a case analysis based on the formula identified by \verb+H4+
will identify only a single case, as the term structure 
\verb+(abs T3 ([x]R x))+ uniquely identifies a single structure for \verb+D2+,
that where the head is $\ofabs$.
We may now use the weakening metatheorem for LF to change the context
for the assumption formula identified by \verb+H6+ and the ones
resulting from the case analysis of the formula identified by \verb+H4+ to 
\verb+(G, n1:tm, n2:of n1 T3)+.
Noting that this context must satisfy the context schema \verb+ctx+ if
\verb+G+ satisfies it, we have the ingredients in place to utilize the
induction hypothesis, \ie\ the assumption formula identified by
\verb+IH+, to be able to add the formula 
\begin{verbatim}
   {G, n1:tm, n2:of n1 T3 |- D1 n5 n4 n3 n2 n1 n : eq T4 T5}
\end{verbatim}
to the collection of assumption formulas; note that the conclusion
formula for the sequent would have also become
\begin{verbatim}
   exists D3, {G |- D3 : eq (arr T3 T4) (arr T3 T5)}
\end{verbatim}
as a result of the case analysis.
Analyzing this assumption formula will produce
a single case in which \verb+D1+ is bound to \verb+(refl T5)+ and
\verb+T4+ is replaced by \verb+T5+ throughout the sequent.
This branch of the proof can then be completed by instantiating the
existential quantifier in the goal formula with the term \verb+(refl (arr T3 T5))+
and using the assumption formulas that indicate that \verb+T3+ and
\verb+T5+ must have the type \sty\ to conclude that \verb+(arr T3 T5)+
must also be similarly typed. 

The second subgoal, the one corresponding to the application case, 
will now be presented in elaborated form.
We will elide a discussion of this case because it does not illustrate
any capabilities beyond those already exhibited in the consideration
of the abstraction case.
Instead we will jump to the final subgoal which is manifest in the
following proof state: 
\begin{verbatim}
Vars: D2:o -> o -> o, T2:o -> o -> o, T1:o -> o -> o
Nominals: n1:o, n:o
Contexts: G{}:c[(n:tm, n1:of n (T1 n n1))]
IH: ctx G:c, forall E:o T1:o T2:o D1:o D2:o, 
      {G |- T1 : ty} => {G |- T2 : ty} => {G |- D1 : of E T1}* =>
      {G |- D2 : of E T2} => exists D3:o, {G |- D3 : eq T1 T2}
H1:{G |- T1 n n1 : ty}
H2:{G |- T2 n n1 : ty}
H4:{G |- D2 n n1 : of n (T2 n n1)}

==================================
exists D3, {G |- D3 : eq (T1 n n1) (T2 n n1)}
\end{verbatim}
As we can see in this state, a new block has been elaborated in the context 
variable type of {\tt G}.
The nominal constants that are introduced by this elaboration may be
used in the terms that instantiate the eigenvariables \verb+D2+,
\verb+T2+, and \verb+T1+ in the original sequent, a fact that is
realizes here by raising these variables over the nominal constants. 
The key to the proof for this case is that case analysis based on the assumption 
formula \verb+H4+ will only identify a single relevant case: that when \verb+D2+ 
is \verb+n1+ and \verb+T2+ is identified with \verb+T1+.
This is because \verb+n+ represents a unique nominal constant
that cannot be matched with any other term or nominal constant, and a name can have 
at most one binding in \verb+G+.

\subsubsection{Cut Admissibility for the Intuitionistic Sequent Calculus}

The example we consider now is that of proving the admissibility of
cut for an intuitionistic sequent calculus.
In this proof, there is a need to consider the weakening of the
antecedents of sequents.
We propose an encoding of sequents below under which the weakening
rule applied to sequents can be modelled by a weakening applied to LF
typing judgements.
Thus, this example illustrates the use of LF metatheorems encoded in
Adelfa in directly realizing reasoning steps in informal proofs.

We will actually consider only a fragment of the intuitionistic
sequent calculus for propositional logic in this illustration.
In particular, this fragment includes only the logical constant $\top$
and the connectives for conjunction and implication.
An LF signature that encodes this fragment appears below.

\begin{center}
\begin{tabular}{lcl}
$\propty:\type$ & \hspace{1.5in} & $\toptm:\propty$\\[3pt]
$\hypty:\imp{\propty}{\type}$ & &$\imptm:\imp{\propty}{\imp{\propty}{\propty}}$\\
$\concty:\imp{\propty}{\type}$ & & $\andtm:\imp{\propty}{\imp{\propty}{\propty}}$\\[5pt]
\end{tabular}

\begin{tabular}{l}
$\inittm:\typedpi{A}{\propty}{\typedpi{D}{\hypty\app A}{\concty\app A}}$\\[3pt]
$\toprtm:\concty\app\toptm$\\[3pt]
$\andltm:\typedpi{A}{\propty}{\typedpi{B}{\propty}{\typedpi{C}{\propty}{\typedpi{D1}{(\typedpi{x}{\hypty\app A}{\typedpi{y}{\hypty\app B}{\concty\app C}})}{}}}}$\\
\qquad\quad $\typedpi{D2}{\hypty\app (\andtm\app A\app B)}{\concty\app C}$\\[3pt]
$\andrtm:\typedpi{A}{\propty}{\typedpi{B}{\propty}}{\typedpi{D_1}{\concty\app A}{\typedpi{D_2}{\concty\app B}{\concty\app (\andtm\app A\app B)}}}$\\[3pt]
$\impltm:\typedpi{A}{\propty}{\typedpi{B}{\propty}{\typedpi{C}{\propty}{\typedpi{D_1}{\concty\app A}{\typedpi{D_2}{(\typedpi{x}{\hypty\app B}{\concty\app C})}{}}}}}$\\
\qquad\quad $\typedpi{D_3}{\hypty\app (\imptm\app A\app B)}{\concty\app C}$\\[3pt]
$\imprtm:\typedpi{A}{\propty}{\typedpi{B}{\propty}{\typedpi{D}{(\typedpi{x}{\hypty\app A}{\concty\app B})}{\concty\app (\imptm\app A\app B)}}}$
\end{tabular}
\end{center}

\noindent The propositions of this calculus are encoded as terms of
type $\propty$.
The two LF type constructors $\hypty$ and $\concty$ that take terms of
type $\propty$ as arguments are used to identify propositions that
are hypotheses and conclusions of a sequent respectively.
More specifically, a sequent of the form $\intseq{A_1,\ldots A_n}{B}$
is represented under this encoding by the LF typing judgement
\[
\lfcheck{x_1:\hypty~\overline{A_1},\ldots,x_n:\hypty~\overline{A_n}}{c}{\concty~\overline{B}},
\]
where $\overline{A}$ denotes the encoding of the proposition $A$.

The contexts relevant to this example can be characterized
by the following schema declaration:
\begin{verbatim}
Schema cutctx := {A : o}(x : hyp A)
\end{verbatim}
Relative to the given signature and schema declaration, the
admissibility of cut is captured by the following formula:
\begin{tabbing}
\quad\=\quad\=\kill
\>$\fctx{G}{cutctx}{\fall{a:\oty}{\fall{b:\oty}{\fall{d_1:\oty}{\fall{d_2:\imp{\oty}{\oty}}{}}}}}$\\
\>\>$\fimp{\fatm{\emptyce}{a:\propty}}
     {\fimp{\fatm{G}{d_1:\concty\app a}}
     {\fimp{\fatm{G}{\lflam{x}d_2~x:\typedpi{x}{\hypty~a}{\concty\app b}}}{\fexists{d:\oty}{\fatm{G}{d:\concty~b}}}}}$
\end{tabbing}
Note that in the third antecedent the term variable $d_2$ is permitted to depend on
the cut formula $a$ through the explicit dependency on the bound variable $x$, reflecting the fact that this proof can use the ``cut formula.''

The informal proof that directs our development uses a nested
induction.
The first induction is on the structure of the cut formula,
expressed by the formula $\fatm{\emptyce}{a:\propty}$.
The second induction is on the height of the derivation that uses the
cut formula, expressed by
$\fatm{G}{\lflam{x}d_2~x:\typedpi{x}{\hypty~a}{\concty\app b}}$.
The full proof is based on considering the different possibilities for
the last rule in the proof with the cut formula included in the
assumption set.
In this illustration, we consider only the case where this is the rule
that introduces an implication on the right side of the sequent.
In the Adelfa proof, the indicated analysis is reflected in the case
analysis of the formula 
$\fatm{G,n:\hypty\app a}{d_2\app n:\concty\app b}$, which is derived
from $\fatm{G}{\lflam{x}d_2~x:\typedpi{x}{\hypty~a}{\concty\app b}}$,
and the case to be considered is that where the head of the term
$(d_2\app n)$ is chosen to be $\imprtm$.  

One of the characteristics of LF typing judgements is that the
assignments in typing contexts are ordered so as to reflect possible 
dependencies.
However, in the object system under consideration, the ordering of
propositions on the left of the sequent arrow is unimportant.
This is evident in our encoding by the fact that there are no
means to represent dependencies in the kind of contexts in
consideration.
Thus, we may use the permutation metatheorem to realize reordering in
the context in a proof development that requires this.
We will need to employ this idea in order to ensure that the
context structure has the form $(G, n:\hypty\app a)$ as is needed to
invoke the induction hypothesis.

To get to the details of the case under consideration, we note that
the cut formula is preserved through the concluding rule in the
derivation. The informal argument then uses the induction hypothesis
to obtain proofs of the premises of the rule but where the cut formula
is left out of the left side of the sequents.  These proofs are then
combined to get a proof for the concluding sequent, again with the cut
formula left out from the premises, using the same rule to introduce
an implication on the right. It is this argument that we want to mimic
in the Adelfa development.

The proof state at the start of the case that is the focus of our
discussion is depicted below:

\begin{verbatim}
Vars: d:o -> o -> o, b1:o, b2:o, d1:o, a:o
Nominals: n1:o, n:o
Contexts: G:cutctx[]
IH:ctx G:cutctx. forall a:o, forall b:o, forall d1:o, forall d2:o -> o,
        {a : proptm}* => {G |- d1 : conc a} => 
            {G, n:hyp a |- d2 n : conc b} => exists d:o, {G |- d : conc b}
IH1:ctx G:cutctx. forall a:o, forall b:o, forall d:o1, forall d2:o -> o,
        {a : proptm}@ => {G |- d1 : conc a} =>
            {G, n:hyp a |- d2 n : conc b}** => exists d:o, {G |- d : conc b}
H1:{a : proptm}@
H2:{G |- d1 : conc a}
H3:{G, n:hyp a |- impR b1 b2 ([x] d n x) : conc (imp b1 b2)}@@
H4:{G, n:hyp a |- b1 : proptm}**
H5:{G, n:hyp a |- b2 : proptm}**
H6:{G, n:hyp a, n1:hyp b1 |- d n n1 : conc b2}**
==================================
exists d, {G |- d : conc (imp b1 b2)}
\end{verbatim}

\noindent For simplicity, we have elided the other subgoals that remain
in this display. 
Our objective in this case is to add to the assumptions the formula
\verb+{G, n1:hyp b1 |- D : conc b2}+ for some term \verb+D+; from this, we can
easily construct a term that we can use the instantiate the
existential quantifier in the goal formula to conclude the proof.
We would like to use the inductive hypothesis identified by \verb+IH1+
towards this end. 
To be able to do this, we would need to extend the context expression
in the assumption formula identified by {\tt H2} with a new binding of
type {\tt (hyp b1)}.
We can do this using a tactic command that encodes the weakening
metatheorem for LF.
In actually carrying out this step, we would additionally
use the strengthening metatheorem for LF to conclude that the type
\verb+(hyp b)+ is  well-formed in the relevant context because of the
assumption identified by \verb+H4+ and the fact that \verb+b1+
cannot depend on \verb+n+.  
We must also rearrange
the type assignments in the context in the assumption formula
\verb+H6+ so that the assignment corresponding to the cut formula
appears at the end.
This can be realized using the tactic command that encodes the
permutation metatheorem in LF, a use of which will be valid for
the reasons identified in the earlier discussion. 

The end result of applying these reasoning steps is a state of the
following form:
\begin{verbatim}
Vars: d:o -> o -> o, b1:o, b2:o, d1:o, a:o
Nominals: n1:o, n:o
Contexts: G:cutctx[]
IH:ctx G:cutctx. forall a:o, forall b:o, forall d1:o, forall d2:o -> o,
        {a : proptm}* => {G |- d1 : conc a} => 
            {G, n:hyp a |- d2 n : conc b} => exists d:o, {G |- d : conc b}
IH1:ctx G:cutctx. forall a:o, forall b:o, forall d1:o, forall d2:o -> o,
        {a : proptm}@ => {G |- d1 : conc a} =>
            {G, n:hyp a |- d2 n : conc b}** => exists d:o, {G |- d : conc b}
H1:{a : proptm}@
H2:{G |- d1 : conc a}
H3:{G, n:hyp a |- impR b1 b2 ([x] d n x) : conc (imp b1 b2)}@@
H4:{G, n:hyp a |- b1 : proptm}**
H5:{G, n:hyp a |- b2 : proptm}**
H6:{G, n:hyp a, n1:hyp b1 |- d n n1 : conc b2}**
H7:{G |- b1 : proptm}**
H8:{G, n1:hyp b1 |- d1 : conc a}
H9:{G, n1:hyp b1, n:hyp a |- d n n1 : conc b2}**
==================================
exists d, {G |- d : conc (imp b1 b2)}
\end{verbatim}
We can now use the induction hypothesis \verb+IH1+ to the assumption
formulas \verb+H1+, \verb+H8+, and \verb+H9+
to introduce a new term variable \verb+d'+ and the assumption formula
\verb+{G, n1:hyp b1 |- d' n1 : conc b2}+, from which we can complete
the proof as previously indicated.

\subsubsection{A Disjunctive Property for Natural Numbers}\label{sssec:natnum}

The formulas that we have considered up to this point have had a
``function'' structure: for every term satisfying certain typing
constraints they have posited the existence of a term satisfying
another typing constraint.
Towards exhibiting the flexibility of the logic underlying Adelfa, we
now consider an example of a theorem that \emph{does not} adhere to
this structure. 
This example is based on the following signature that encodes natural
numbers and the even and odd properties pertaining to these numbers.
\begin{center}
\begin{tabular}{lclcl}
$nat : \type$ & \qquad & $even : \imp{nat}{\type}$ & \qquad & $odd : \imp{nat}{\type}$\\[3pt]
$z : nat$ & & $\mbox{\sl e-z} : even~z$ & & $\mbox{\sl e-o} : \typedpi{N}{nat}{\imp{even~N}{odd~(s~N)}}$\\
$s : \imp{nat}{nat}$ & & $\mbox{\sl o-e} : \typedpi{N}{nat}{\imp{odd~N}{even~(s~N)}}$ & &
\end{tabular}
\end{center}
\noindent Given this signature, we can express the property that every natural
number is either even or odd: 
\begin{center}
$\fall{N:\oty}{\fimp{\fatm{\emptyce}{N:nat}}{\for{(\fexists{D:\oty}{\fatm{\emptyce}{D:even~N}})}{(\fexists{D:\oty}{\fatm{\emptyce}{D:odd~N}})}}}.$
\end{center}
Following the obvious informal proof, this formula can be proved by induction
on the height of the derivation of the LF typing judgement
$\lfcheck{\emptyce}{N}{nat}$.
The application of the induction hypothesis yields two possible cases
and in each case we pick a disjunct in the conclusion that we can show
to be true.

\section{Conclusion}\label{sec:conclusion}

This paper has described the Adelfa system for reasoning about specifications
written in LF.
Adelfa is based on the logic \logic\ in which the atomic formulas represent
typing judgements in LF and where quantification is permitted over
terms and over contexts that are characterized by context schemas.
We have sketched a proof system for this logic and have demonstrated
the effectiveness of Adelfa, which realizes this proof system, through
a few examples.
We have developed formalizations beyond the ones discussed here, such
as type preservation for the STLC and subtyping for F$_{<:}$; the
latter is a problem from the PoPLMark
collection~\cite{aydemir05tphols}.

While this work is not unique in its objectives, it differs
from other developments in how it attempts to achieve them.
One prominent approach in this context is that adopted by what we
refer to as the ``Twelf family.''
The first exemplar of this approach is the Twelf
system~\cite{Pfenning99cade} that allows properties of interest to be 
characterized by types and the validity of such properties to be
demonstrated by exhibiting the totality as functions of inhabitants of
these types.
This approach has achieved much success but it is also limited by the
fact that properties that can be reasoned about must be encodable by a
function type, i.e. they must be formulas of the
$\forall...\exists...$ form.
By contrast, the logic underlying Adelfa has a much more flexible
structure, allowing us, for example, to represent properties with a
disjunctive structure as we have seen in Section~\ref{sssec:natnum};
the reader might want to consult the encoding of this property
provided in the wiki page of the Twelf project that covers output
factoring to understand more fully the content of this
observation~\cite{twelf.website}. 
Another virtue of our approach is that we are able to exhibit an
explicit proof for properties at the end of a development.
This drawback is mitigated to an extent in the Twelf context by the
presence of the logic $M_2^+$~\cite{schurmann00phd} that provides a
formal counterpart to the approach to reasoning embodied in the Twelf
system~\cite{wang13lfmtp}.

The Twelf approach differs in another way from the approach described
here: it does not provide an explicit means for quantifying over
contexts.
It is possible to parameterize a development by a context description,
but it is one fixed context that then permeates the development.
As an example, it is not possible to express the strengthening lemma 
pertaining to equality of types in the STLC that we discussed in
Section~\ref{ssec:example}.
The Beluga system~\cite{pientka10ijcar}, which otherwise shares the
characteristics of the Twelf system, alleviates this problem by using
a richer version of type theory that allows for an explicit treatment
of contexts as its basis~\cite{nanevski08tocl}.

An alternative approach, that is more in alignment with what
we have described here, is one that uses a translation of LF specifications 
into predicate logic to then be reasoned about using the Abella
system~\cite{southern14fsttcs}. 
This approach has several auxiliary benefits deriving from the
expressiveness of the logic underlying Abella~\cite{gacek11ic}; for
example, it is possible to define relations between contexts, to treat
binding notions explicitly in the reasoning process through the
$\nabla$-quantifier, and to use inductive (and co-inductive)
definitions in the reasoning logic.
In future work, we would like to examine how to derive some of these
benefits in the context of Adelfa as well.
There are, however, also some drawbacks to the translation-based
approach. 
One of the problems is that it is based on a ``theorem'' about the
translation~\cite{felty90cade} that we know to be false for the
version of LF that it pertains to.\footnote{The
  theorem in question states that a (closed) LF typing judgement
  is derivable if and only if the translated form of the judgement is
  derivable in the relevant predicate logic. 
This is true in the forward direction.
However, in the version of LF that the theorem is about, the
translation loses typing information from terms and, hence, the
inverse translation is ambiguous.
Thus, it is only a weaker conclusion that can be drawn, that there is
\emph{some} LF judgement that is derivable if the formula in predicate
logic is derivable; there is no guarantee that this is the same LF
judgement that we started out with.}
Another issue is that proof steps that can be taken in Abella are
governed by the logic underlying that system and these allow for many
more possibilities than are sensible in the LF context.
Moreover, it is not immediately clear how the reasoning steps that are
natural with LF specifications translate into macro Abella steps
relative to the translation. 
In this respect, one interesting outcome of the work on the logic
underlying the Adelfa system might be an understanding of how one might
build within Abella (or other related proof assistants) a targeted
capability for reasoning about LF specifications.
This also raises another interesting possibility that is worthy of
attention, that of providing an alternative justification for the
proof system for \logic\ based on the translation.

\section*{Acknowledgements}

This paper is based upon work supported by the National Science
Foundation under Grant No. CCF-1617771.
Any opinions, findings, and conclusions or recommendations expressed
in this material are those of the authors and do not necessarily
reflect the views of the National Science Foundation.

\bibliographystyle{eptcs}
\bibliography{../../references/master}
\end{document}